\documentclass[12pt,preprint]{aastex}

\usepackage{graphicx}

\begin{document}

\title{Evidence for the Missing Baryons in the Angular Correlation
of the Diffuse X-ray Background}

\author{M. Galeazzi\altaffilmark{1}, A. Gupta, and E. Ursino\altaffilmark{2}}
\affil{Physics Department, University of Miami, Coral Gables, FL 33124}
\altaffiltext{1}{Corresponding author, galeazzi@physics.miami.edu}
\altaffiltext{1}{Current address: Universit\`{a} di Roma 3, Rome, Italy}

\begin{abstract}
The amount of detected baryons in the local Universe is at least a factor 
of two smaller than measured at high redshift. It is believed that 
a significant fraction of the baryons in the current Universe is 
``hiding'' in a hot filamentary structure filling the intergalactic space, 
the Warm-Hot Intergalactic Medium ($WHIM$).  
We found evidence of the missing baryons in the $WHIM$ by detecting their 
signature on the angular correlation of diffuse X-ray emission with 
the XMM-Newton satellite. Our result indicates that $(12\pm 5)$\% of 
the total diffuse X-ray emission in the energy range 0.4-0.6~keV is due to 
intergalactic filaments. 
The statistical significance of our detection is several sigmas 
($\chi ^2>136$ $N=19$). The error bar in the X-ray flux is dominated, instead, 
by cosmic variation and model uncertainties.
\end{abstract}


\maketitle

\section{Introduction}

Several methods have been used to measure the quantity of baryons in the 
Universe. Measurements at high redshift all point to about 4\% of the 
matter-energy density of the Universe to be in the form of baryons, 
while the rest consists of dark matter and dark energy \cite{rauch, weinberg,
burles,kirkman,bennet}. 
In contrast, the amount of baryons measured in the local Universe in 
the form of stars, galaxies, groups, and clusters is less than 
2\% \cite{fukugita}. 

Hydrodynamic simulations tracking the 
evolution of the Universe suggest that much of the ``missing'' material 
in the local Universe lies in a hot filamentary gas filling the 
intergalactic space, the Warm-Hot Intergalactic Medium ($WHIM$). 
Theoretical estimates indicate that the $WHIM$ filaments, have 
typical density 20 to 1000 times the 
average baryon density of the universe and temperature greater than 
$5\times 10^5$~K \cite{cen99,borgani,cen06}. 
At these temperatures and densities the baryons 
are in the form of highly ionized plasma, making them invisible to all 
but low energy X-ray and UV observations. Recent UV 
observations \cite{danforth} 
confirm the presence of ``warm'' gas in the intergalactic medium that 
should account for half of the missing baryons. However, the ``hot'' core 
of the filaments can only be detected in the soft X-ray band, mostly 
through excitation lines of highly ionized heavy elements, such as 
O{\tiny VII}, O{\tiny VIII}, Fe{\tiny XVII}, C{\tiny VI}, and N{\tiny VII}
with energy between 0.2 and 0.7~keV.
 
Different approaches have been proposed to detect and study the missing 
baryons in X-rays. Absorption features can be spotted along the line of 
sight of distant bright sources. Evidence of the filament existence
comes from an absorption measurement made in the direction of the blazar 
Markarian 421 in a period of maximum brightness \cite{nicastro05,nicastro05b}, 
however such 
data are still considered controversial \cite{kaastra,williams,rasmussen}. 
Unfortunately, with the 
capability of current X-ray satellites, Markarian 421 is practically 
the only source sufficiently bright to be used as beacon for 
absorption measurements. 

Another option is to identify red-shifted X-ray emission lines emitted 
by the filaments \cite{ursino}, but the emission 
is very weak and must be detected on top of a larger background from 
galactic emission and unresolved point sources. Dedicated missions are on 
the drawing board for the study of the WHIM  (e.g., den Herder et al. - 2007), 
but, if approved, they are still several years 
away. With the capability of current missions, searching for individual
filaments
in emission is impractical, except in a few specific directions where filaments
are expected. Clusters of galaxies form in the higher density regions 
(or nodes) of the filamentary structure and are connected by lower density 
filaments identified with the $WHIM$. 
Following the cluster network could therefore point to the location of the
filaments. The complexity of the network and the distance between clusters 
makes this approach very difficult, except in few cases where 
clusters are sufficiently close together. Recently, an observation between 
the the clusters Abell 222 and Abell 223 has found evidence of excess 
emission attributed to a filament connecting them \cite{werner}.

A different way to detect the $WHIM$ in emission is to use a statistical
approach. The plasma filaments are expected to have a characteristic angular 
structure that can be identified and studied using the angular 
autocorrelation function \cite{ursino,croft}.  Hydrodynamic simulations have been 
used to predict the angular distribution of soft X-rays emitted by the 
filaments. The results show that the filaments should have typical 
angular scales of a few arcminutes or smaller, leaving a clear signature 
in the angular distribution of their X-ray emission \cite{ursino}. The difficulty 
in detecting such signature lies in the fact that the filament emission 
only represents a fraction of the total diffuse X-ray emission, which 
is dominated by unresolved point sources and diffuse galactic emission, 
such as Local Bubble and Galactic Halo \cite{galeazzi,mushotzky}. 
Data from the ROSAT 
All-Sky Survey ($RASS$) and ROSAT pointed observations have been used to 
calculate the angular autocorrelation function on scales greater than 
2 arcmin, but the results were dominated by emission from unresolved 
point sources \cite{soltan,kuntz}. The angular autocorrelation function has also 
been used to quantify the possible detection of a filament using the 
ROSAT $PSPC$ detector \cite{scharf}.  

The XMM-Newton satellite, launched in 1999, is currently the best 
platform to investigate the angular distribution of X-rays emitted by 
the missing baryons. The CCD detectors aboard XMM-Newton ($PN$ and $MOS$) 
have a field of view of about 30' and an angular resolution of about 
14~arcsec, covering the range of angular scales of interest. Moreover, 
the angular resolution 
of the detectors allows good source identification, strongly reducing 
the effect of unresolved point sources. The detectors also maintain 
a reasonable response down to about 0.2~keV, covering the energy range 
of interest.

\section{Data Reduction and Analysis}

For our investigation we combined several observations available in the 
XMM-Newton public archive, including two ``proprietary'' observations 
designed for this investigation. We used data from 11 observations, 
corresponding to 6 different targets. The observation characteristics 
are summarized in Table~\ref{table1}. The choice of the targets was based on 
several considerations. To limit the effect of absorption due to the 
neutral hydrogen and contamination from galactic emission we used targets 
at least $30^o$ above the galactic plane and with a neutral hydrogen 
column density smaller than $2\times 10^{20}$~atoms~cm$^{-2}$. 
The only exception is our proprietary observation in the direction of 
the neutral hydrogen cloud $MBM20$, which, in conjunction with the 
linked observation of a very low neutral hydrogen density region 
nearby, the Eridanus Hole, has been used as ``control'' observation. 
We also limited the investigation to targets with at least 80,000~s of 
observing time and sufficiently distant from any bright point source.

To characterize the angular correlation of the diffuse X-ray background 
we used the angular autocorrelation function ($AcF$) $w(\theta)$ which, 
in the following form, has also been used for the characterization of 
the diffuse X-ray emission \cite{soltan}:
\begin{equation}
w(\theta)=\frac{<I(n)I(n')>}{<I>^2}-1,
\end{equation}
where $I(n)$ is the intensity in the direction $n$, $I(n)I(n')$ is the
product of intensities with angular separation $\theta$, and $<...>$ 
denotes the expectation values of the corresponding quantities.
A practical recipe to calculate the $AcF$ on a single image, using a 
weighting of the intensity based on exposure time, is \cite{ursino}: 
\begin{equation}
w(\theta)=\frac{\sum_{i=1,N_\theta}\sum_{j=i+1,N_\theta}
\Big(R_i-\overline{R}\Big)\Big(R_j-\overline{R} \Big)\sqrt{s_is_j}}
{\sum_{i=1,N_\theta}\sum_{j=i+1,N_\theta}\sqrt{s_is_j}}\cdot
\frac{1}{\overline{R}^2}.
\end{equation}
where the sum is over the $N_{\theta}$ pairs of pixels separated by $\theta$, 
$R$ is the count rate in those pixels, $s$ is the exposure time for 
those pixels, used as statistical weight, and $\overline{R}$ denotes 
the average count rate.

Even when no bright objects are present in an X-ray field, the image 
contains the contribution from the Diffuse X-ray background and the 
detector intrinsic background. Moreover, in the soft X-ray band 
(below 1~keV) the Diffuse X-ray Background does not have a unique 
source, but is the overposition of contributions from 5 different 
sources, which are, in order of distance from the Earth 
(e.g., Galeazzi et al. - 2007):
\begin{itemize}
\item {\it Solar Wind Charge Exchange.} When the highly ionized solar 
wind interacts with neutral gas in our atmosphere and in the 
interplanetary medium, an electron may jump from the neutral to the 
ion. The electron then cascades to the lower energy level of the ion, 
emitting soft X-rays in the process.
\item {\it Local Bubble.} It is believed that the sun lies in a cavity 
of the interstellar medium extending about 100~pc in all directions, 
filled with 1,000,000~K plasma. The plasma is highly ionized, emitting 
radiation in the soft X-ray band. 
\item {\it Galactic halo.} Our galaxy is surrounded by a ``halo'' of 
high temperature plasma, probably due to intergalactic gas falling into 
the gravitational well of the Milky Way. The temperature of such gas is 
believed to be higher than that of the Local Bubble, emitting 
in the soft X-ray band at slightly higher energy.
\item {\it Intergalactic gas.} The focus of this investigation. 
\item {\it Unresolved point sources.} A significant contribution to 
the diffuse X-ray background comes from point sources (mainly $AGN$s) 
that are too faint to be individually resolved by the X-ray telescope.
\end{itemize}

When calculating the $AcF$ of a blank field in the sky, the result is 
therefore the combination of all mentioned contribution, according to 
the expression:
\begin{equation}
n_{tot}^2 w(\theta)=\sum_{i=1,N} n_i ^2 w_i (\theta),
\end{equation}
where $n_{tot}$ and $n_i$ represent the total flux and the fluxes of 
each individual component, $w(\theta)$ is the total $AcF$, and 
$w_i (\theta)$ are the $AcF$ of each individual component. Solar 
Wind Charge Exchange, Local Bubble, and Galactic Halo have no correlation 
in the angular scale investigated and $w_i (\theta)=0$, however their 
flux will contribute to $n_{tot}$.
The equation can therefore be rewritten as:
\begin{equation}
n_{tot}^2 w(\theta)=n_B ^2 w_B (\theta)+n_P ^2 w_P (\theta)+n_W ^2 w_W (\theta),
\label{eq_AcF}
\end{equation}
where the index $B$ indicates detector background, $P$ point sources, 
and $W$ the $WHIM$.

\section{Contamination from Other Sources}

We discuss here the steps taken to properly assess and remove
all sources of possible contamination and their effect on our 
result.

\subsection{Unresolved point sources.} 

Unresolved point sources, 
including active galactic nuclei ($AGN$), blazars, galaxies, galaxy 
clusters and groups, and stars are expected to have a characteristic 
angular signature in the same angular range of filaments, possibly 
shadowing the signature of the missing baryons. The total contribution 
of point sources to the diffuse X-ray emission has been well 
characterized in the energy band of interest for our investigation 
\cite{mushotzky,giacconi,mccammon,gandhi}. 
Using the angular resolution of XMM-Newton and the long 
exposure time of the observations, we have bee able to identify and 
remove the brightest sources, corresponding to $(68\pm 5)$\% of the 
total point source flux. The source identification is done using 
XMM-Science Analysis Software point source detection algorithm that 
iteratively identifies point sources $3\sigma$ above threshold using 
images both in the 0.5-2~keV and 0.4-0.6~keV energy ranges. 

The removal of the unresolved point sources contribution is the most critical 
part of the analysis. The total flux from point sources has been 
estimated by many authors. 
For our work, we used a conservative approach based on data from 
McCammon et al. (2002). The McCammon et al. (2002) paper reports a high spectral 
resolution investigation of the diffuse X-ray background. The 
high spectral resolution allows an easy identification of emission lines 
at redshift zero, and we used the flux outside these lines as an upper limit 
to the point source emission. For the 
evaluation of the shape of the $AcF$, we assumed instead that the shape 
does not change between resolved and unresolved point sources. 
While by definition unresolved point sources are weaker than the 
identified ones, potentially belonging to a different population, it 
has been shown that the contribution of weak sources to the $AcF$ is 
equal or smaller than that of bright sources \cite{giacconi}. 
We also used a combined XMM-Newton - Chandra list of identified point 
sources in the Hubble Deep Field N  target. Using the Chandra 
catalogue improves by a factor of 30 the threshold of identified 
point sources (from $7.2\times 10^{-16}$~ergs~s$^{-1}$~cm$^{-2}$ to 
$2.3\times 10^{-17}$~ergs~s$^{-1}$~cm$^{-2}$) and does not reduce 
the value of the measured AcF. 
A control ``experiment'' using the value of the $AcF$ in a different
energy range fully confirms the validity of our procedure and is 
discussed in the results section.

\subsection{Instrumental background}

The instrumental background is expected to produce a uniform ``counts'' 
image; however, when we divide the uniform "counts" image by the 
non uniform exposure map to extract a ``flux'' image, we artificially 
generate an angular correlation due to the detector background. 
In our analysis the background flux is measured using spectral 
information, while its $AcF$ is 
calculated by using a flat image divided by the exposure map.
The procedure we used to evaluate the instrumental background in 
XMM-Newton data has been described in details and tested 
in Galeazzi et al. (2007).

\subsection{Absorption from the neutral hydrogen} 

The neutral hydrogen column density may not be uniform across 
each individual field of view, potentially introducing an artificial 
angular correlation due to its X-ray absorption. We used $IRAS$ 100 $\mu m$
data to generate neutral hydrogen maps of the fields used in this 
investigation and to evaluate the $AcF$ signal due to the neutral hydrogen 
distribution. We verified that the effect is negligible. 

\subsection{Control observations} 

We used our proprietary observations of MBM20 and the Eridanus hole to 
verify that all spurious ``local'' effects, such as Local Bubble 
non-uniformity and instrumental contributions, have been properly 
identified and removed and the correlation signal that we see is 
due to the extragalactic emission. The average neutral hydrogen 
column density of MBM20 is $15.9\times 10^{20}$~cm$^{-2}$, absorbing 
approximately 75\% of the non local X-ray flux in the energy of 
interest (including from the missing baryons), and 
$0.86\times 10^{20}$~cm$^{-2}$ in the Eridanus hole observation, 
absorbing only about 8\% of the non local X-ray flux \cite{galeazzi}. 
The two targets are very close to each other (about $2^o$ apart) and 
therefore should not be affected by any significant spatial variation 
in the magnitude of the $WHIM$ $AcF$ (``Cosmic Variance''). 

\section{Results}

To maximize the signal from the filaments compared to other 
sources, we focused on a narrow energy band between 0.4 and 
0.6~keV, where the O{\tiny VII} and O{\tiny VIII} emission 
from filaments at redshift up to about 0.5 are concentrated. 
Figure~1 shows the average $AcF$ of all the observations 
excluding MBM20. The contributions due to unresolved point 
sources, instrumental background, and neutral hydrogen 
absorption have already been removed. Each individual observation 
shows a strong non-zero correlation for angles below about 6 arcmin, 
each with a significance of several sigmas ($\chi ^2>136$ for every 
field, $N=19$ - See Fig.~2 and Table~\ref{table2}). 
The error bars in Fig.~1 are a combination of the statistical
errors from each target (Fig.~2) and a statistical error due to 
Cosmic variance,  i.e., variation between targets, 
as predicted by simulations. The latter is calculated simply as 
the standard deviation in the $AcF$ values from the different targets.
The error bars are dominated by cosmic variance, and do not represent 
the statistical limit of the measurement.  

Figure~3 shows the calculated AcF for the control target MBM20. 
In this case the AcF is compatible with zero ($\chi ^2=16$, $n=19$). 
The Eridanus hole $AcF$ in the same energy range, in contrast, is 
similar to the average low neutral hydrogen density $AcF$ and 
significantly different from zero for all angles below 6 arcmin 
($\chi ^2=136$, $n=19$). The result confirms the effectiveness of 
our procedure at removing all ``local'' sources of contamination.
To verify the validity of our procedure at removing the contribution 
from point sources and confirm our result we repeated the same 
procedure used to calculate the $AcF$ for the Eridanus Hole field 
in the energy range 0.7-0.9~keV. The new energy range is very 
close to the original one and the contribution from point sources 
is expected to be very similar. However, the new range is just above 
the energy of most lines that are emitted by the intergalactic 
filaments and no significant contribution from the $WHIM$ is expected. 
The Eridanus Hole $AcF$ from the new energy band is also shown in 
Fig.~3. The result is compatible with zero ($\chi ^2=25$, $n=18$) 
and is incompatible with the AcF from the same target for the energy 
band 0.4-0.6~keV ($\chi ^2=88$, $n=18$). This result clearly supports 
the effectiveness of our procedure and the validity of the claimed 
detection in the lower energy band.

We also compared the experimental result with the expected $AcF$ signal 
derived from our simulations of the filament emission (Fig.~1). 
After subtracting the contribution from detector background and point 
sources Eq.~\ref{eq_AcF} can be written as:
\begin{equation}
n_{tot}^2 w(\theta)=n_W ^2 w_W (\theta),
\end{equation}
where $n_{tot}$ includes, in this case, the flux from all the components of 
the diffuse X-ray background except point sources that have already 
been removed. To evaluate the X-ray flux from the $WHIM$ ($n_W$), we used 
the total X-ray flux in the range 0.4-0.6~keV minus the flux from point 
sources as an estimate of $n_{tot}$, the measured $AcF$ from 
Fig.~1 for $w(\theta )$, and the results from our simulations of 
the X-ray emission from the $WHIM$ (e.g., Ursino \& Galeazzi - 2006, 
Ursino et al. - 2008 in preparation, see Fig.~1) 
for $w_W (\theta)$. 
The errors from experimental data and simulations, and the spread in the 
simulations due to the different models have been combined to estimate the final error.
We found that the fraction of X-rays due to the missing baryons in the 
0.4-0.6 keV band is $(12\pm5)$\% of the total diffuse X-ray emission (including
point sources) in the same band. This is in agreement with theoretical predictions 
and simulations \cite{ursino,croft,phillips}.

\section{Conclusions}

We studied the angular distribution of the diffuse X-ray emission from
6 targets available in the XMM-Newton public archive.
After removing all knows sources of contamination we found a clear 
signal that we attribute to emission from the Warm-Hot Intergalactic Medium. 
A comparison with simulations indicate that, in the energy range 0.4-0.6~keV,
the WHIM emission correspond to $(12\pm 5)$\% of the total diffuse X-ray 
emission. The statistical significance of our detection is several sigmas 
($\chi ^2>136$ $N=19$). The error bar in the X-ray flux is dominated, instead, 
by cosmic variation and model uncertainties.

\acknowledgments

This work has been supported in part by the National Aeronautic and Space 
Administration. The authors would like to thank Dan McCammon, Frits Paerels, 
Enzo Branchini, Stefano Borgani, Lauro Moscardini, and Mauro Roncarelli.

\clearpage
\begin{figure}
\begin{center}
\includegraphics[height=8cm]{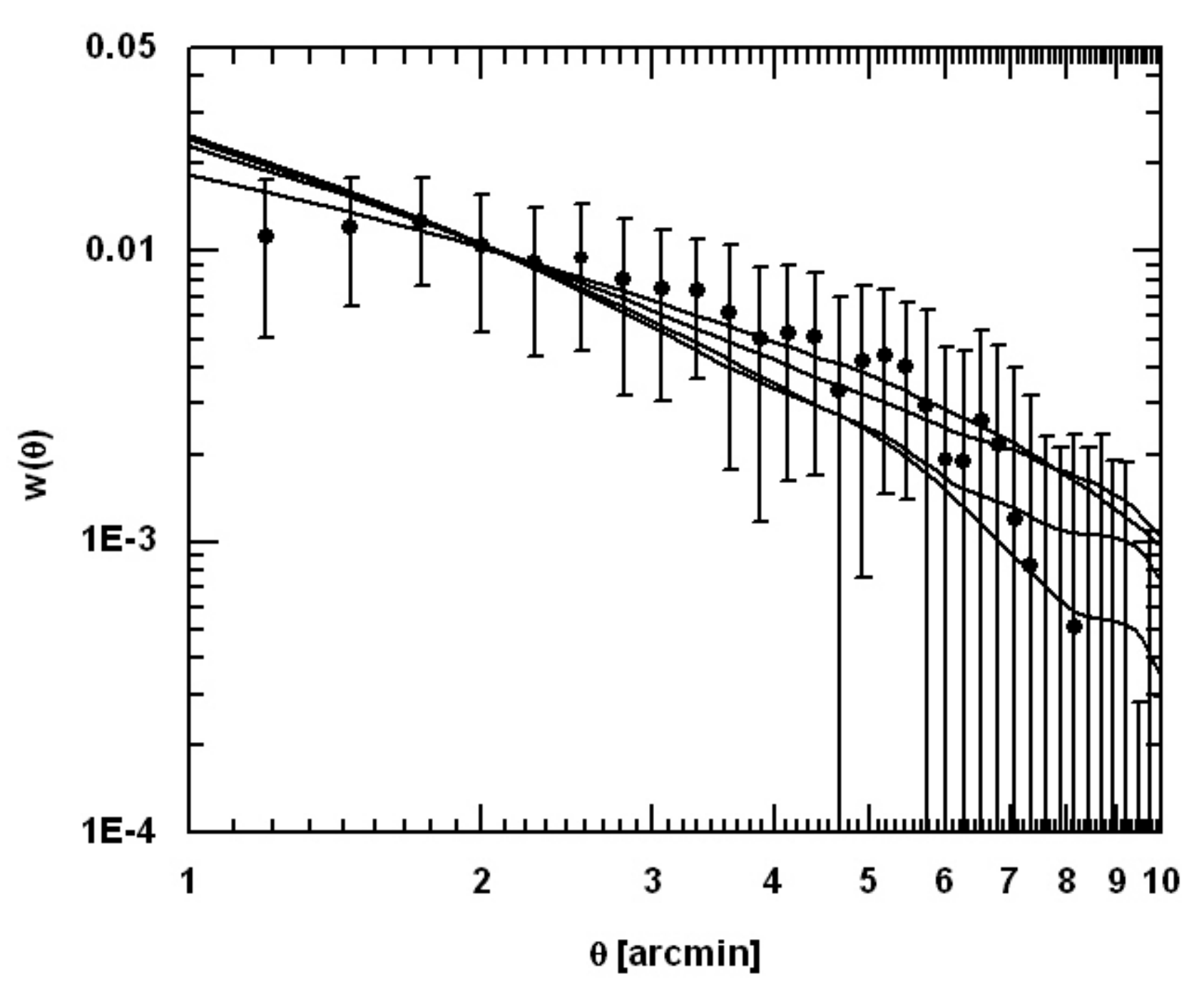}
\caption{Average $AcF$ for all the targets used in this investigation 
except MBM20. The error bars are a combination of statistical error and 
cosmic variation, and are dominated by the latter. The curves represent 
the best fit to the experimental data points using the output of four 
different models. For each model the only free parameter in the fit is 
a constant scaling factor dependent on the fraction of X-rays due to 
missing baryons compared to the total diffuse X-ray emission.}
\end{center}
\label{fig1}
\end{figure}

\clearpage
\begin{figure}
\begin{center}
\includegraphics[height=8cm]{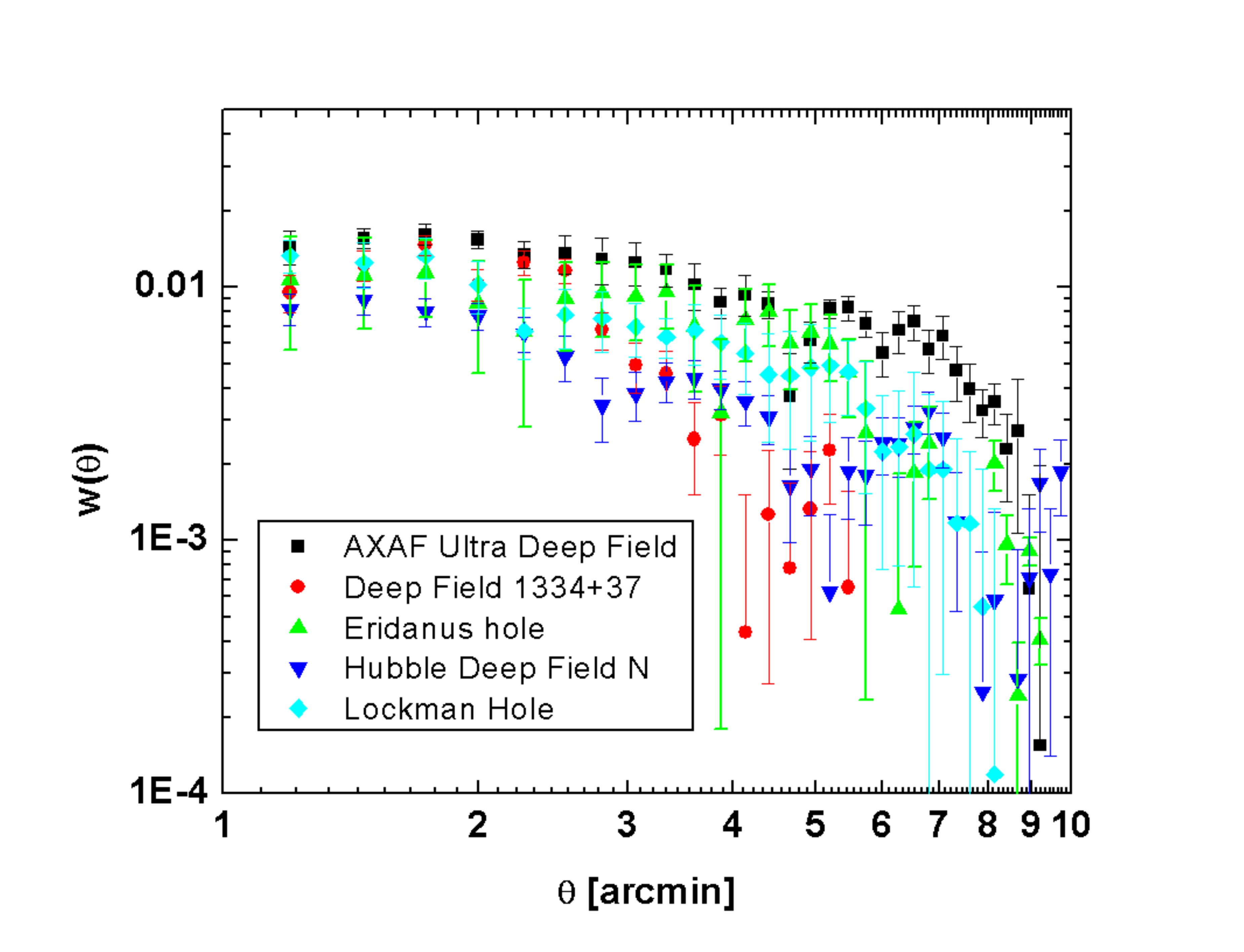}
\caption{Measured $AcF$ for each individual target used in this investigation.
While, as expected, there is a significant difference in the value of the $AcF$
between targets (cosmic variation), the value of the $AcF$ for each target
is significantly different from 0 below 6'.}
\end{center}
\label{fig2}
\end{figure}

\clearpage
\begin{figure}
\begin{center}
\includegraphics[height=8cm]{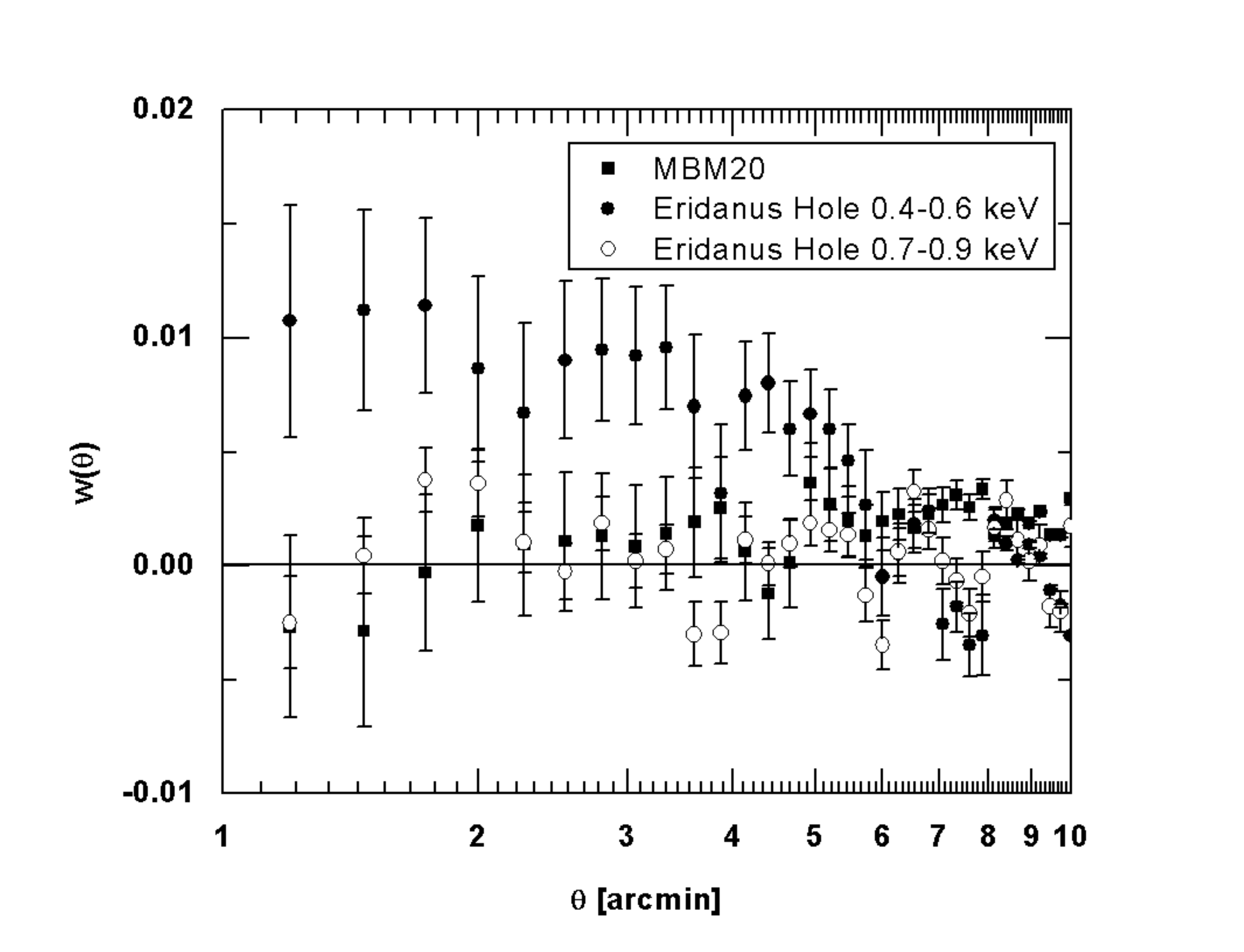}
\caption{Calculated $AcF$ for the two control targets, MBM20 (squares) 
and the Eridanus hole (full circles), in the energy band 0.4-0.6~keV, 
and for the Eridanus hole in the 0.7-0.9~keV energy band (empty circles). 
The MBM20 $AcF$ is compatible with zero, confirming the accuracy of 
our procedure to remove the local contribution to the $AcF$. The 
Eridanus hole AcF for the energy band 0.7-0.9~keV is also compatible 
with zero, confirming the accuracy of our procedure to remove the 
point source contribution to the $AcF$. The Eridanus Hole $AcF$ in 
the energy band 0.4-0.6~keV is instead statistically different from 
zero, due to the contribution from the $WHIM$.}
\end{center}
\label{fig3}
\end{figure}

\clearpage
\begin{deluxetable}{lcccc}
\tablewidth{0pc}
\tablecaption{Summary of the targets used in this investigation.}
\tablehead{
\colhead{Target}           & \colhead{l}      &
\colhead{b}          & \colhead{NH($10^{20}$ $cm^{-2}$)} & Exposure (s)}
\startdata
Lockman Hole & 149 16 48.1 & 53 08 45.9 & 0.7 & 628,945\\
Hubble Deep Field N & 125 53 31.3 & 54 48 52.4 & 1.4 & 131,751\\
Deep Field 1334+37 & 85 37 22.5 & 75 55 16.4 & 0.8 & 165,989\\
Eridanus Hole & 213 25 17.9 & -39 04 25.6 & 0.86 & 50,230\\
AXAF Ultra deep F & 223 34 36 & -54 26 33.3 & 1.0 & 431,618\\
MBM20 & 211 23 53.2 & 36 32 41.8 & 15.9 & 31,419 \\
\enddata
\label{table1}
\end{deluxetable}

\clearpage
\begin{deluxetable}{lcc}
\tablewidth{0pc}
\tablecaption{Statistical significance of the $AcF$ for each target
used in the investigation, compared with a zero value hypothesis.}
\tablehead{
\colhead{Target}           & \colhead{$\chi ^2$}      &
\colhead{Degrees of Freedom}}
\startdata
Lockman Hole & 283 & 19 \\
Hubble Deep Field N & 1152 & 19 \\
Deep Field 1334+37 & 540 & 19 \\
Eridanus Hole & 514 & 19 \\
AXAF Ultra deep F & 136 & 19 \\
\enddata
\label{table2}
\end{deluxetable}

\end{document}